\documentclass[10pt,a4paper]{article}
\usepackage{amsmath}
\usepackage{amssymb}

\def\beq{\begin{equation}}
\def\eeq{\end{equation}}

\def\be{\begin{displaymath}}
\def\ee{\end{displaymath}}

\newtheorem{thm}{Theorem}

\title{
Rigidity, Functional Equations and the Calogero-Moser Model
}

\author{H. W. Braden\thanks{E-mail:hwb@ed.ac.uk}\\
\normalsize
\em Department of Mathematics and Statistics,\\
\normalsize
\em The University of Edinburgh, \\
\normalsize
\em Edinburgh, UK \\
}

\date{}

\begin{document}

\renewcommand{\thepage}{}

\maketitle
\vskip-9.5cm
\hskip10.4cm
\sf solv-int/0005046 \rm
\vskip8.8cm

\begin{abstract}
Suppose we have a natural Hamiltonian $H$ of $n$ particles on the line, 
centre of mass momentum $P$ and a further independent quantity $Q$, cubic 
in the momenta. If these are each $S_{n}$ invariant and mutually Poisson 
commute we have the Calogero-Moser system with potential
$V=\frac{1}{6}\sum\limits_{i\neq j}\wp\left( q_{i}-q_{j}\right) +const$.

\end{abstract}



\section{Introduction}
The following note deals with many particle Hamiltonian systems on the
line and their integrability. 
Although such systems arise in many physical settings and have been
extensively studied
there still is no simple way to determine their integrability or otherwise.
General arguments \cite{FO} tell us that many particle Hamiltonian systems
for sufficiently repulsive potentials are integrable, yet there appear few
direct methods of actually solving for such systems. 
The integrable systems we can actually solve seem to form a very privileged
class. The result presented here sheds some light on this state of affairs.
We will follow a less well known route to the study of integrable systems,
that employing functional equations.

Here we address the following question: What $S_n$ invariant, natural 
Hamiltonian systems of $n$-particles on the line and conserved
centre of mass momentum admit a third independent,
 $S_n$ invariant, mutually Poisson 
commuting quantity, cubic in the momenta? (The precise statement and explanation
of these terms will be given below.) Our answer is somewhat surprising.
This data characterises the $a_n$ Calogero-Moser systems. This is the 
``rigidity" of our title. Although no restrictions were placed on further
Poisson commuting invariants we arrive at a system for which sufficient exist
to yield complete integrability. The mixture of symmetry and polynomial
momentum is powerful. Such natural requirements and our result go some way in
explaining the ubiquity of this class of models.
The situation is somewhat reminiscent of
the original work of Ruijsenaars and Schneider
\cite{RS} who, when demanding certain commutation properties, discovered
a class of Hamiltonian systems that proved to be integrable. It is also 
analogous to what
one encounters with W and related algebras, where a few commutation
relations specify the whole structure. Indeed, the connections between Conformal
Field Theory and these models may mean this is more than analogy \cite{mat, fv}.

There are obvious generalisations to this work which will be taken up in the
discussion. Before turning to the statement and proof of the result (given
in the following two sections) it is
perhaps worth making some general remarks on connections between integrable
systems and functional equations.
Functional equations have of course a long and interesting history
in connection with mathematical physics and touch upon many branches
of mathematics \cite{Acza, Acz}.
They have arisen in the context of completely integrable systems in
several different ways.
We have already mentioned the mentioned the work of Ruijsenaars and Schneider.
Hietarinta similarly derived a functional equation when seeking a second
quartic integral for two particle systems on the line \cite{Hiet}.
A further way in which they arise is by assuming an
ansatz for a Lax pair, the consistency of the Lax pair yielding functional
and algebraic constraints. In this manner Calogero discovered the elliptic
Calogero-Moser model \cite{Ca2} and Bruschi and Calogero constructed
Lax pairs for the Ruijsenaars models \cite{BCa, BCb}. The functional equations
found by this route appear \cite{BB1} as particular examples of 
\begin{equation*}
\phi_1(x+y)=
\frac{
\begin{vmatrix}\phi_2(x)&\phi_2(y)\cr\phi_3(x)&\phi_3(y)\end{vmatrix}
     }
     {
\begin{vmatrix}
\phi_4(x)&\phi_4(y)\cr\phi_5(x)&\phi_5(y)
\end{vmatrix}
     }.
\end{equation*}
The general analytic solution of this has been given by Braden and
Buchstaber \cite{BB2}.
Interestingly, Novikov's school have shown that the Hirzebruch genera associated
with the index theorems of known elliptic operators arise as solutions of 
functional equations which are particular examples of this.
The string inspired Witten index was shown by Ochanine to 
be described by Hirzebruch's construction where now the elliptic function
solutions were important \cite{HBJ}.
A similar approach based upon an ansatz and consequent functional
equations was used by Inozemtsev \cite{Inoz} to construct generalisations
of the Calogero-Moser models, while  in \cite{BMMM} this route was used
to construct new solutions to the WDVV equations.
Various functional equations have also arisen when studying the properties of
wave-functions for associated quantum integrable problems. Gutkin found several
functional relations by requiring a nondiffractive  potential \cite{Gu} while
Calogero \cite{cal} and Sutherland \cite{suth, Su2} obtained
functional relations by seeking factorizable ground-state wave-functions.
A recurring equation  in this latter approach is
\begin{equation*}
\begin{vmatrix}
1              & 1              & 1              \\
f(x)           & g(y)           & h(z)           \\
f\sp{\prime}(x)& g\sp{\prime}(y)& h\sp{\prime}(z)\\
\end{vmatrix}
=0,\quad\quad x+y+z=0.
\label{reddetdiff}
\end{equation*}
This finds general solution in \cite{bp, BBS}. In our present work we will
make use of the particular case of this equation \cite{bp, BBS}:
\begin{thm}
Let f be a three-times differentiable function satisfying the functional
equation
\begin{equation}
\begin{vmatrix}
1              & 1              & 1              \\
f(x)           & f(y)           & f(z)           \\
f\sp{\prime}(x)& f\sp{\prime}(y)& f\sp{\prime}(z)\\
\end{vmatrix}
=0,\quad\quad  x+y+z=0.
\label{detdiff}
\end{equation}
Up to the manifest invariance
\begin{equation*}
f(x)\rightarrow \alpha f(\delta x)+\beta,
\label{equiv}
\end{equation*}
the solutions of (\ref{detdiff}) are one of $f(x)=\wp(x+d)$,
$f(x)=e\sp{x}$ or $f(x)=x$.
Here $\wp$ is the Weierstrass $\wp$-function and 
 $3 d$ is a lattice point of the $\wp$-function.
\end{thm}

Perhaps one reason for the underlying connection between integrability and
functional equations is that fact that Baker-Akhiezer functions obey such
relations. Such connections between integrable functional equations and
algebraic geometry have been studied by Buchstaber and Krichever \cite{BK}
and Dubrovin, Fokas and Santini \cite{DFS}. Whatever, these connections between
functional equations and complete integrability warrant further
investigation.

\section{The Result}
The result discovered is the following:
\begin{thm}
Let $H$ and $P$ be the (natural) Hamiltonian and centre of mass momentum  
\begin{equation}
H= \frac{1}{2} \sum\limits_{i=1}^{n}p_{i}^{2}+V,\qquad
P= \sum\limits_{i=1}^{n}p_{i}.
\label{HP}
\end{equation}
Denote by $Q$ an independent third order quantity
\begin{equation}
Q=\sum\limits_{i=1}^{n}p_{i}^{3}+
\frac{1}{6}\sum\limits_{i\ne j\ne k}d_{ijk}p_i p_j p_k
+\sum\limits_{i\ne j}d_{ij}p_i^2  p_j+
\frac{1}{2}\sum\limits_{ij}a_{ij}p_{i}p_{j}+\sum_ib_{i}p_{i} +c.
\label{Q}
\end{equation}
If these are $S_{n}$ invariant and Poisson commute,
\begin{equation*}
\left\{ P,H\right\} =\left\{ P,Q\right\} =\left\{ Q,H\right\} =0,
\end{equation*}
then
$V=\frac{1}{6}\sum\limits_{i\neq j}\wp\left( q_{i}-q_{j}\right) +const$
and we have the Calogero-Moser system.
\end{thm}

Some explanatory remarks are in order.
Here $S_{n}$ invariance means that for any coefficient 
$\alpha_{ij}(q_1,q_2,\ldots,q_n)$ in the expansions above we have
$\alpha_{\sigma(i) \sigma(j)}(q_{\sigma(1)},q_{\sigma(2)},\ldots,q_{\sigma(n)})$
for all $\sigma \in S_{n}$. In particular
$V( q_{1},q_{2},\ldots,q_{n}) =
 V( q_{\sigma (1)},q_{\sigma (2)},\ldots ,q_{\sigma(n)})$
for all $\sigma \in S_{n}$. We remark that had we begun with 
particles of possibly different particle masses,
 $H=\frac{1}{2} \sum\limits_{i=1}^{n}m_i p_{i}^{2}+V$,
the effect of $S_{n}$ invariance is such as to require these masses
to be the same. Thus we are assuming the $S_{n}$ invariant Hamiltonian of the
introduction.
Finally, by ``an independent third order quantity" $Q$ we mean one functionally
independent of $H$ and $P$ and for which one cannot obtain an invariant of
lower degree by subtracting multiples of $P^3$ and $PH$.
We are not dealing with quadratic conserved quantities here.

\section{The Proof}
Our proof has five steps.
We begin by noting that 
the Poisson commutativity $\left\{ Q,H\right\} =0$ yields
(with $\left\{ q_{i},p_{j}\right\} =\delta _{ij}$)
\begin{equation}
\begin{split}
0 =&
\frac{1}{6}\sum_l \sum\limits_{i\ne j\ne k}\left(\partial_l d_{ijk}\right)
p_i p_j p_k p_l
+\sum_l \sum\limits_{i\ne j}\left(\partial_l d_{ij}\right) p_i^2  p_j p_l 
+ \sum\limits_{i,j,l}\left(\partial_l a_{ij}\right)p_i p_j p_l\\
&+\sum\limits_{i,j}\left(\partial_{i}b_j\right) p_i p_j
-3\sum\limits_{i}p_{i}^{2}\left(\partial_{i}V\right)
-\frac{1}{2}\sum\limits_{i\ne j\ne k}d_{ijk}\left(\partial_k V \right)p_i p_j \\
&-\sum\limits_{i\ne j}d_{ij}\left(2\left(\partial_{i}V\right) p_i p_j+
      \left(\partial_{j}V\right) p_i\sp2\right)
-\sum\limits_{i,j}a_{ij} \left(\partial_{i}V\right) p_{j}
+\sum\limits_{i}\left(\partial_{i}c\right) p_{i}-
\sum\limits_{i}b_i\partial_{i}V .
\end{split}
\label{PC}
\end{equation}

The steps then are:
\begin{enumerate}

\item First
we show that the $d_{ijk}$ and $d_{ij}$ terms in (\ref{Q})
may be taken to be zero.

Having made this simplification we then focus on the terms  remaining
in (\ref{PC}) independent and quadratic in the momenta:
\begin{eqnarray}
 \partial _{j}b_{i}+\partial _{i}b_{j} &=&0, \qquad i\ne j,  \label{eq1} \\
 \partial _{i}b_{i}-3\partial_iV &=&0,  \label{eq2} \\
 \sum b_{i}\partial _{i}V &=&0.  \label{eq3}
\end{eqnarray}
\item Second, using (\ref{eq1},\ref{eq2}) we show that $b_j$ may be written
in the form
\begin{equation}
b_{j} =\sum\limits_{i\neq j}W\left( q_{i}-q_{j}\right) +U\left(
q_{j}\right) , \label{eqbu}
\end{equation}
where $W$ is an even function.

\item Third, using $\left\{ P,Q\right\} =0$, we may set $U=0$.

\item Fourth, that we may rewrite (\ref{eq3}) in the form
\begin{equation}
0=\sum\limits_{i< j< k}\left| 
\begin{array}{ccc}
1 & 1 & 1 \\
W(q_i-q_j) & W(q_j-q_k) & W(q_k-q_i) \\
W'(q_i-q_j) & W'(q_j-q_k) & W'(q_k-q_i)
\end{array}
\right|.
\label{suma}
\end{equation}
\item Finally we argue that each term in the sum (\ref{suma}) itself vanishes
and so we arrive at an equation of the form (\ref{detdiff}). 
The result then follows simply.

\end{enumerate}

\noindent{\bf {Step 1.}}
We begin by focusing on the terms in (\ref{PC}) quartic in the momenta.
For $l$ different from $i,j,k$ we see that  $ \partial_l d_{ijk}=0$,
and so $d_{ijk}=d_{ijk}(q_i, q_j, q_k)$. Further, from the coefficients of
$p_i\sp3 p_j$, $p_i\sp2p_j\sp2$ and $p_i\sp2p_j p_k$ (for $i,j,k$ distinct)
respectively, we find
\begin{equation}
\partial_i d_{ij}=0,\qquad
\partial_j d_{ij}=0,\qquad
\partial_j d_{ik}+\partial_k d_{ij}+\partial_i d_{ijk}=0.
\label{step1}
\end{equation}
The first and third of these together show $\partial_i\sp2 d_{ijk}=0$
and so $d_{ijk}$ is at most linear in $q_i$. By symmetry 
$$
d_{ijk}=\alpha q_i q_j q_k +\beta(q_i q_j+q_j q_k+q_k q_i)+\gamma (
q_i +q_j+q_k) +\delta.
$$
Now using $\left\{ P,Q\right\} =0$ shows $\alpha=\beta=\gamma=0$.
Thus $d_{ijk}$ is a constant. This fact, together with the second and
third equations of (\ref{step1}), shows $\partial_k\sp2 d_{ij}=0$. Therefore
$d_{ij}$ is at most linear in $q_k$ (for $k\ne i,j$). The first two equations
of (\ref{step1}) show $d_{ij}$ is independent of $q_i$ and $q_j$.
Now a similar argument
employing $\left\{ P,Q\right\} =0$ yields $d_{ij}$ to be constant.
By subtracting appropriate multiples of $P^3$ and $PH$ we may 
then remove the $d$ terms from $Q$. Our assumption
of independence means that the leading term of $Q$ does not vanish when
doing this. Thus (after such a subtraction and a possible rescaling) we may 
set the $d$ terms in
$Q$ to be zero. Henceforth we will assume this simplification has been made.

\noindent{\bf {Step 2.}}
Suppose $i,j, k$ are distinct. Then from (\ref{eq1}) we obtain 
($\partial _{ij}=\partial _{i}\partial _{j}$ etc.)
$$
\partial _{jk}b_{i}+\partial _{ik}b_{j}=0, \qquad
\partial _{jk}b_{i}+\partial _{ij}b_{k}=0.
$$
Taking the difference of these we see 
$\partial _{i}\left(\partial _{k}b_{j}-\partial _{j}b_{k}\right)=0$
and so 
$$-\partial _{k}b_{j}+\partial _{j}b_{k}=2F\left(q_j,q_k\right).$$
Combining this with $\partial _{k}b_{j}+\partial _{j}b_{k}=0$ we
obtain
$$
\partial _{j}b_{k}=F\left( q_j,q_k\right) =-F\left( q_k,q_j\right)
=-\partial _{k}b_{j}.
$$
We wish to further restrict the form of $F$.
If we apply $\partial _i$ to (\ref{eq1}) and then use (\ref{eq2}) we
see
$$
-\partial _{i}\partial _{i}b_{j}=\partial _{i}\partial _{j}b_{i}=
3\partial _{i}\partial _{j}V=\partial_{j}\partial _{i}b_{j}
$$
and so
$$
\left( \partial _{i}+\partial _{j}\right) \partial _{i}b_{j}=0.
$$
Therefore
\begin{equation}
\partial _{i}b_{j}=F\left( q_{i}-q_{j}\right),\qquad F(x)=-F(-x).
\label{eq5}
\end{equation}
Upon integrating we obtain (\ref{eqbu}) where $W^{\prime }(x)=F(x)$ and
$W$ is an even function.
(In principle upon integrating the odd function $F$ we obtain a function
$\widetilde W$ where $\widetilde W^{\prime }(x)=F(x)$ and
$\widetilde W(x)=\widetilde W(-x)+\tilde c$. {\it A priori} we cannot
argue that the integration constant $\tilde c$ vanishes if
$\widetilde W(0)$ is not defined, as happens for singular potentials.
However setting $W(x)=\frac{1}{2}\left(\widetilde W(x)+\widetilde W(-x)\right)$
again yields (\ref{eqbu}) up to a constant, which at this stage may be
incorporated into the arbitrary function $U$.)
We have employed the $S_n$ symmetry throughout this step to
identify each of the possibly different functions $F$, $W$ and $U$ arising
from each pair as the same.

\noindent{\bf {Step 3.}}
Now upon employing $\left\{ Q,P\right\} =0$ we see 
$\sum\limits_{i=1}^{n}\partial _{i}b_{j}=0$. Using (\ref{eqbu})  we
deduce that $\partial _{i}U(q_{i})=0$ and so $U(q_{i})$ is a constant.
Such a constant may be removed altogether by subtracting an appropriate
multiple of $P$ from $Q$, or simply incorporated into a redefinition
of $W(x)$. Whatever, we may take $U=0$.  Then
\begin{equation}
b_{i}^{2}=\sum\limits_{j\neq i}W^{2}\left( q_{j}-q_{i}\right)
+2\sum\limits_{j\neq k\neq i}W\left( q_{j}-q_{i}\right) W\left(
q_{k}-q_{i}\right) .
\label{eqb2}
\end{equation}

\noindent{\bf {Step 4.}}
Now employing (\ref{eq2}, \ref{eq3}) we see $0=\sum_i \partial _{i}b_{i}^{2}$.
Using (\ref{eqb2}) we obtain
\begin{equation*}
\partial _{i}b_{i}\sp2= -2\sum\limits_{j\neq i}W\left( q_{j}-q_{i}\right)
F\left( q_{j}-q_{i}\right) +2\sum\limits_{j\neq k\neq i}\partial _{i}\left(
W\left( q_{j}-q_{i}\right) W\left( q_{k}-q_{i}\right) \right)
\end{equation*}
When we sum this expression over $i$ the first term will vanish
using oddness and evenness properties. Thus we arrive at
$$ 0=\sum\limits_{i\neq j\neq k}\partial _{i}\left( W\left(
q_{j}-q_{i}\right) W\left( q_{k}-q_{i}\right) \right). $$

Define $A_{ijk}$ by
\begin{equation*}
A_{ijk}=\partial _{i}\left( W_{ji}W_{ki}\right) +\partial _{j}\left(
W_{ij}W_{kj}\right) +\partial _{k}\left( W_{ik}W_{jk}\right) =\left| 
\begin{array}{ccc}
1 & 1 & 1 \\ 
W_{ij} & W_{jk} & W_{ki} \\ 
F_{ij} & F_{jk} & F_{ki}
\end{array}
\right|, \end{equation*}
where we use the shorthand $W_{ij}=W\left(q_{i}-q_{j}\right)$.
Then from the functional form of $W$ we know
\begin{equation}
A_{ijk}= A_{jik}=A_{jki}=
\Psi \left( q_{i}-q_{j},q_{j}-q_{k},q_{k}-q_{i}\right)\label{fform}
\end{equation}
and is fully symmetric in $i,j,k$. Thus
\begin{equation}
0=\sum\limits_{i< j< k}A_{ijk}=\sum\limits_{i< j< k}\left|
\begin{array}{ccc}
1 & 1 & 1 \\
W(q_i-q_j) & W(q_j-q_k) & W(q_k-q_i) \\
W'(q_i-q_j) & W'(q_j-q_k) & W'(q_k-q_i)
\end{array}
\right|.
\label{sumb}
\end{equation}
which is equation (\ref{suma}).

\noindent{\bf {Step 5.}}
We now wish to argue that $A_{ijk}=0$. 
If we apply $\partial _{ijk}$ to
(\ref{sumb}) we find that
$$\partial _{ijk}A_{ijk}=0,$$
as only one term in the sum depends on $i,j,k$. Thus 
$\partial _{jk}A_{ijk}$ is independent of $q_i$, and consequently due to the
functional form (\ref{fform}) it must be a function of $q_{j}-q_{k}$ only.
Therefore we must have
$$\partial _{jk}A_{ijk}=B\left( q_{j}-q_{k}\right),$$
and so, after integration and use of symmetry,
$$A_{ijk}=E\left( q_{i}-q_{j}\right) +E\left( q_{j}-q_{k}\right)+
E\left( q_{k}-q_{i}\right)$$
(where $E(x)=E(-x)$ and $E^{\prime \prime }(x)=-B(x)$). We may therefore rewrite
(\ref{sumb}) as
\begin{equation}
0=\sum\limits_{i<j}E\left( q_{i}-q_{j}\right).
\label{sume}
\end{equation}
Taking the partial derivative $\partial _{ij}$ of this expression
then gives $\partial _{ij}E\left( q_{i}-q_{j}\right)=0$, as only this term 
depends on both $i$ and $j$. This, together with the evenness of $E$ tells us
that $E$ is a constant. In conjunction with (\ref{sume}) we deduce $E=0$.
That is $A_{ijk}=0$. Therefore for each distinct triple $i,j,k$
$$
0=\left|
\begin{array}{ccc}
1 & 1 & 1 \\
W(q_i-q_j) & W(q_j-q_k) & W(q_k-q_i) \\
W'(q_i-q_j) & W'(q_j-q_k) & W'(q_k-q_i)
\end{array}
\right|.
$$
But this is none other than (\ref{detdiff}). The even solution of this is
$W(x)=\wp(x)$, up to a constant. 
Finally, using (\ref{eqbu}) and (\ref{eq2}) we obtain the stated conclusion.


\section{Discussion}
Our result may be interpreted as a rigidity theorem for the $a_n$ 
Calogero-Moser system and in part explains this models' ubiquity:
demanding a cubic invariant together with $S_n$ invariance necessitates the
model. A detailed scrutiny of our proof shows several generalisations
possible. A natural generalisation is to replace the $S_n$ invariance with
the invariance of a general Weyl group $W$ and make connection with the
Calogero-Moser models associated to other root systems \cite{OP1, OP2}.
Quite a bit is known about the quantum generalisations in this regard.
Given a commutative  ring $\cal R$ of $W$-invariant, holomorphic, differential
operators, whose highest order terms generate $W$-invariant 
differential operators
with constant coefficients, then the potential term for the Laplacian ${\cal H}$
(the quantum Hamiltonian) has Calogero-Moser potential appropriate to $W$ 
\cite{OS, OOS}. Our result suggests something stronger may be possible: that
the form of the potential may be dictated from just a few elements of $\cal R$.
Taniguchi's results \cite{kt} are indicative of the rigidity of these 
models: if ${\cal H}$ is the quantum Hamiltonian just discussed, and 
${\cal Q}_{1,2}$ are holomorphic (but not {\it a priori} W-invariant)
differential operators of appropriate degrees for which 
$[{\cal Q}_{1,2},{\cal H}]=0$,
then ${\cal Q}_{1,2}\in {\cal R}$ and consequently $[{\cal Q}_1,{\cal Q}_2]=0$.
Interestingly in the present work we have employed a functional equation 
elsewhere encountered in the quantum regime.

A further generalisation of this work
would be to replace the natural Hamiltonian structure
of our theorem with (say) Hamiltonians of Ruijsenaars type. We remark in
passing that there are still several unsolved functional equations 
surrounding this model.
One might also seek to relax the full $S_n$ invariance imposed here.
By so doing this will
allow the Toda models. As shown by Inozemtsev \cite{inoz}, the Toda models
arise as scaling limits of the Calogero-Moser model, the latter being
the ``generic" situation \cite{BB1}. It would be interesting
to understand this in terms of the coadjoint descriptions of these models.

\enlargethispage{.2in}
Though perhaps not obvious, this work arose from trying to understand
models conjectured to be integrable (see for example \cite{BMMM2}).
Given a putative integrable Hamiltonian, what might the invariants look like?
The present work tells us that for $S_n$ invariant systems {\it not} of
Calogero-Moser type we should look for conserved quantities quartic and 
above in the momenta.

\section{Acknowledgements}
I am grateful to A. Mironov, A. Marshakov and A. Morozov together with the
Edinburgh Mathematical Physics group for their comments on this work which
was begun under the support of a Royal Society Joint grant with the FSU.
This work was presented at the Workshop on ``Mathematical Methods of Regular
Dynamics" dedicated to the $150$-th anniversary of Sophie Kovalevski
and I would thank I. Komarov for his remarks.

\end{document}